\documentclass[aip,apl,amsmath,superscriptaddress,amssymb,reprint]{revtex4-2}
\usepackage{chemfig}
\usepackage{amsmath}
\usepackage{amssymb}
\usepackage{color}
\usepackage{times}
\usepackage{epsf}
\usepackage{graphicx}% Include figure files
\usepackage{epsfig}
\usepackage{epstopdf}
\usepackage{dcolumn}% Align table columns on decimal point
\usepackage{bm}% bold math
\usepackage{float}
\usepackage{url}
\usepackage{CJK}
\usepackage{ulem}
\usepackage{makecell}

\def\<{\langle}
\def\>{\rangle}
\def\({\left(}
\def\){\right)}
\def\[{\left[}
\def\]{\right]}

\begin{document}

\title{Thermal percolation and electrical insulation in composite materials with partially metallic coated fillers}

\author{Jinxin Zhong}
\thanks{These authors contribute equally to this work and should be considered co-first authors}
\affiliation{Center for Phononics and Thermal Energy Science, China-EU Joint Lab for Nanophononics,
School of Physics Science and Engineering, Tongji University, 
Shanghai 200092, China}

\author{Qing Xi}
\thanks{These authors contribute equally to this work and should be considered co-first authors}
\affiliation{Research Center for Advanced Science and Technology, The University of Tokyo, Japan}
%author{Tsuneyoshi Nakayama}
%affiliation{Center for Phononics and Thermal Energy Science, China-EU Joint Lab for Nanophononics, School of Physics Science and Engineering, Tongji University, Shanghai 200092, China}
%affiliation{Hokkaido University, Sapporo, Hokkaido 060-0826, Japan}

\author{Jixiong He}
\affiliation{Department of Mechanical and Aerospace Engineering, North Carolina State University, Raleigh, NC 27695, USA}	

\author{Jun Liu}
\email{jliu38@ncsu.edu}
\affiliation{Department of Mechanical and Aerospace Engineering, North Carolina State University, Raleigh, NC 27695, USA}

\author{Jun Zhou}
\email{zhoujunzhou@njnu.edu.cn}
\affiliation{NNU-SULI Thermal Energy Research Center and Center for
  Quantum Transport and Thermal Energy Science, School of Physics and Technology, Nanjing Normal
  University, Nanjing 210023, China}

\date{\today}

\begin{abstract}
	We propose a type of thermal interface materials incorporating insulating nanowires with partially metallic coating in insulating polymer matrix.
	Large thermal conductivity can be obtained due to thermal percolation while the electrical insulation is maintained by controlling $C_{\rm M}\varphi < \varphi_{\rm c}^{\rm e}$ and $\varphi > \varphi_{\rm c}^{\rm th}$, where $\varphi$ is the volume fraction of filler, $C_{\rm M}$ is the metallic coating fraction, $\varphi_{\rm c}^{\rm e}$ and $\varphi_{\rm c}^{\rm th}$ are the electrical and thermal percolation thresholds, respectively.  
	The electrical conductivity of such composite materials can further be regulated by coating configuration. 
	In this regard, we propose the concept of “thermal-percolation electrical-insulation”, providing a guide to design efficient hybrid thermal interface materials.
\end{abstract}
%\PACS{asaa}

\maketitle

Composite materials consisting of polymer matrix and inorganic fillers have been widely used to design the thermal interface materials (TIM) with outstanding performance, especially in the field of electronic devices such as integrated circuits (IC) \cite{bonnet,sun,zheng}.
Various fillers with large thermal conductivity such as Ag nanowire \cite{shi}, carbon nanotube \cite{haggenmueller}, graphite \cite{min,hung}, silicon nitride \cite{zhou}, boron nitride nanotube \cite{jo}, boron nitride nanosheets \cite{wu2020synergistic}, and metal nanofibers \cite{zeng} have been intensively studied. Theoretically, the effective medium theory (EMT) was widely used to calculate the effective thermal conductivity of composite materials \cite{zhao}. 
An effective way to further improve the thermal conductivity of TIM is constructing heat flow networks according to the percolation theory \cite{shi,kargar}. However, EMT is not able to describe the thermal transport once the percolating network is formed, because it neglects the heat exchange among fillers.

Thermal boundary resistance (TBR) plays a crucial role for the performance of TIM. For example, the thermal conductivity of TIMs with electrically conducting fillers (such as silver and graphene) is usually larger than the thermal conductivity of TIMs with electrically insulating fillers (such as $h$-BN) although the thermal conductivity of silver and $h$-BN are comparable. The main reason is the large TBR between dielectric fillers \cite{liuR,wang2020thermal}, while the TBR between metallic fillers is relatively negligible. However, the TIM must be electrical insulator in many applications such as IC packaging. Therefore, it is necessary to propose a strategy to eliminate the large TBR between dielectric fillers. 

In this paper, we propose to use insulating nanowires with partially metallic coating as fillers in TIMs. The TBR between dielectric fillers can be effectively reduced by the contact of coated metals. The coating fraction should be precisely controlled to maintain the electrical insulating of TIM. In other words, the electrical percolating network must not be formed while the thermal percolating network is formed.
We term this concept as “thermal-percolation electrical-insulation” (TP-EI) to achieve high performance TIMs.

\begin{figure}[htp]
	\centering
	\includegraphics[width=0.95\linewidth]{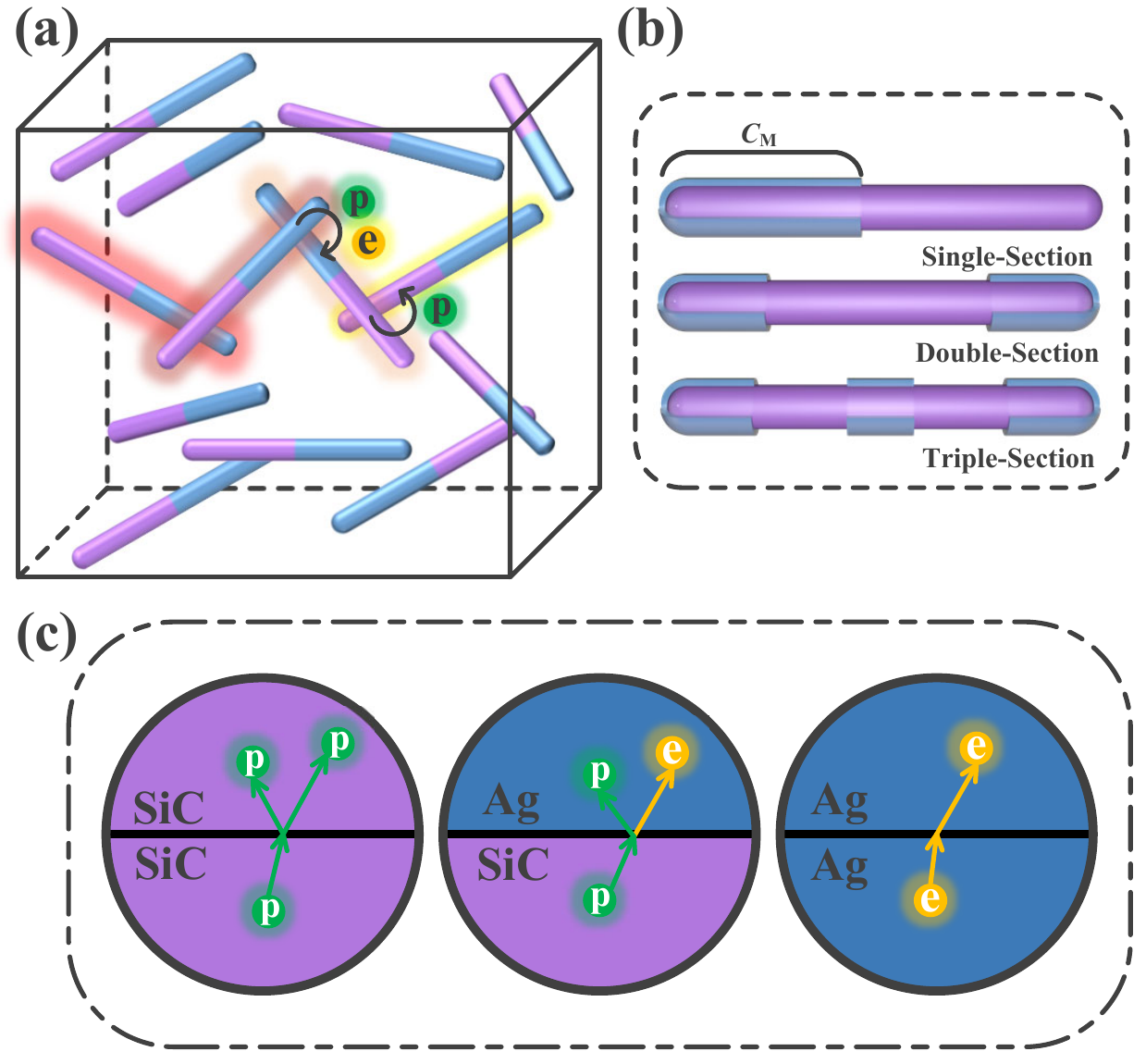}
	\caption{(Color online) (a) Schematic diagram of the concept of TP-EI with partially metallic coating fillers. Here we choose SiC nanowires with Ag coating as an example. Ag is in blue and SiC is in purple. There are three kinds of contacts between fillers: SiC-SiC; Ag-SiC; Ag-Ag. (b) Typical partially coating configurations are used in the calculations: single-section (N=1), double-section (N=2), and triple-section (N=3). (c) Thermal transport across three kinds of interfaces: phonon transport across SiC-SiC interface; coupled electron and phonon transport across SiC-Ag interface; electron transport across Ag-Ag interface. } 
	\label{fig1}
\end{figure}

Figure \ref{fig1} illustrates the TIM with partially metallic coating fillers. The fillers are randomly embedded in a uniform insulating polymer matrix. We choose SiC nanowires with Ag coating to demonstrate the transport process as shown in Fig . \ref{fig1}(a). When the filling fraction of fillers ($\varphi$) is above the structural percolation threshold ($\varphi_{\rm c}$), infinite clusters are formed. Such clusters are essentially the same as heat flow networks. Therefore, the percolation threshold of heat current $\varphi_{\rm c}^{\rm th}=\varphi_{\rm c}$. If a fraction ($C_{\rm M}$) of the surface of SiC nanowires is coated with certain configuration as shown in Fig. \ref{fig1}(b), an electrical percolation network would not form when $C_{\rm  M}\varphi<\varphi_{\rm c}^{\rm e}$. Here the electrical percolation threshold $\varphi_{\rm c}^{\rm e}$ depends on the detailed coating configurations. In short, the TP-EI concept can be realized by choosing $C_{\rm M}\varphi < \varphi_{\rm c}^{\rm e}$ and $\varphi > \varphi_{\rm c}^{\rm th}$ where $0<C_{\rm M}<1$.

There are three kinds of interfaces between contacted fillers as shown in Fig. \ref{fig1}(c): SiC-SiC; Ag-SiC; Ag-Ag. The thermal resistance of these three interfaces can be estimated as follows. 1) For SiC-SiC direct contact, phonons are the heat carriers and the thermal resistance is $R_{\text{SiC-SiC}}/\delta$ where $R_{\text{SiC-SiC}}$ is the TBR of SiC-SiC interface and $\delta$ is the average contact area. 2) For Ag-SiC contact, the thermal resistance is $R_{\text{Ag-SiC}}/S+R_{\text{Ag-SiC}}/\delta$ where $R_{\text{Ag-SiC}}$ is the TBR of Ag-SiC interface. The coating area $S\approx{\pi d l C_{M}/N}$ where $d$ and $l$ are the diameter and length of nanowire, respectively, and $N$ is the number of sections. Both electrons and phonons contribute to the TBR of Ag-SiC interface. Usually, $R_{\text{Ag-SiC}}/\delta$ is dominant because the coating area $S$ is much larger than the contact area $\delta$. (3) For Ag-Ag interface, the overall thermal resistance is $2\times R_{\text{Ag-SiC}}/S+R_{\text{Ag-Ag}}/\delta$. The electrons are the major heat carriers for TBR of Ag-Ag interface. This thermal resistance is negligible compared to previous two contacts because $S\gg \delta$, $R_{\text{Ag-Ag}}\ll R_{\text{Ag-SiC}}$, and $R_{\text{Ag-Ag}}\ll R_{\text{SiC-SiC}}$.

\begin{figure}[htp]
	\centering
	\includegraphics[width=1 \linewidth]{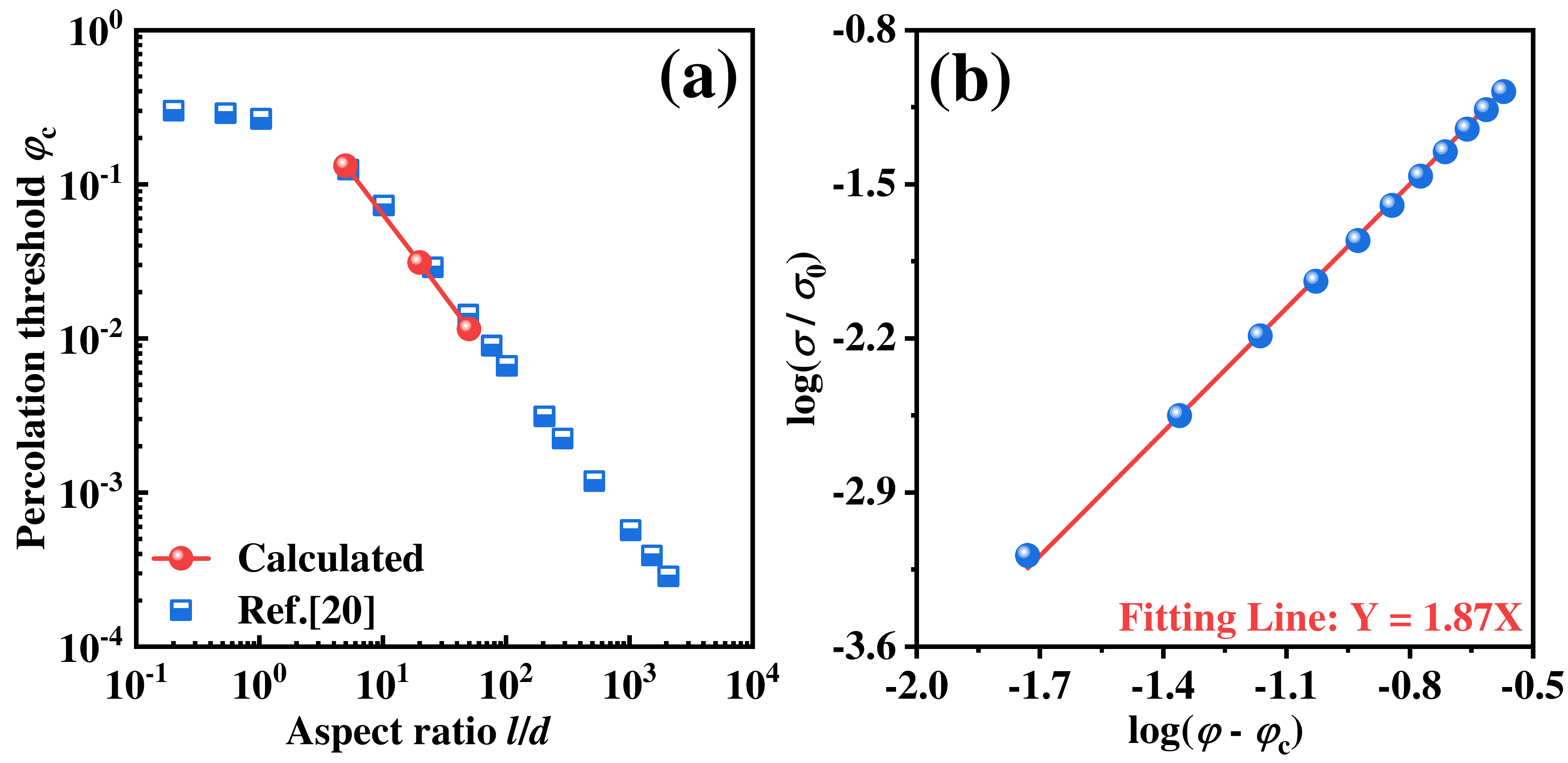}
	\caption{(Color online) Validation: (a) percolation threshold $\varphi_{\rm c}$ versus the aspect ratio $l/d$. (b) log$\sigma/\sigma_{0}$ as a variation of log($\varphi-\varphi_{\rm c}$). $\sigma\propto(\varphi-\varphi_{\rm c})^{t}$ with $t=1.87$ is used to fit the simulation results.} 
	\label{fig2}
\end{figure}

We now use numerical calculations to exemplify the TP-EI concept. The percolation behavior of composite materials has been widely reported \cite{chen,Stankovich,vigolo,nassira} and our numerical simulation method mainly follows Ref.\cite{liu}. We consider that cylindrical SiC nanowires with length $l$ and diameter $d$ are randomly dispersed in a unit cube of width $L$. The thickness of coating $w$ is much thinner than the diameter of nanowires, for example, $w=0.01d$. For the sake of simplicity, the nanowires with $l/L=0.4$ are applied in three-dimensional (3D) Monte Carlo generator. The soft-core nanowires are used in our calculations. It should be pointed out that the difference between hard-core and soft-core is not significant for large aspect ratio in the calculation \cite{foygel}. We first test our calculation when the $C_{\rm M}=1$, i. e. the nanowires are completely coated. Considering an electron current flows along the X-direction, the nanowires that exceed left plane $(x=0)$ and right plane $x>L$ are cut off, and the periodic boundary conditions are applied in the Y- and Z-directions \cite{song}. Furthermore, it is reasonable to assume Ohmic contacts between connected metal coatings because one can efficiently suppress the contact resistance between metal coatings by means of sintering or pressing during the sample preparation \cite{choi}. Applying the voltage bias on the left and right planes, the electrical current can be obtained by adopting a resistor network algorithm and Kirchhoff’s current law for given volume fractions \cite{kirkpatrick}, while if without an electrical percolating network, the electrical current should be zero. The calculation results are shown in Fig.\ref{fig2}. The calculated percolation thresholds by this model agree with other literature \cite{foygel}. And the electrical conductivity $\sigma$ in such system obeys a universal power scaling law,\cite{nakayama,foygel}

\begin{equation}
\sigma = \sigma_{0}(\varphi-\varphi_{\rm c})^{t},
\label{EC}
\end{equation}
where $\sigma_{0}$ is the pre-factor, $\varphi$ is the volume fraction of nanowires, $\varphi_{\rm c}$ is the percolation thresholds, and $t$ is the exponent factor. 
The exponent factor $t$ is obtained to be 1.87 by fitting the simulated data in the log-log plot in Fig.\ref{fig2}(b). 
This agrees with the typical value of 1.8 reported in other literature \cite{lightsey,shenogina,abeles}.

\begin{figure*}[htp]
	\centering
	\includegraphics[width=0.9 \linewidth]{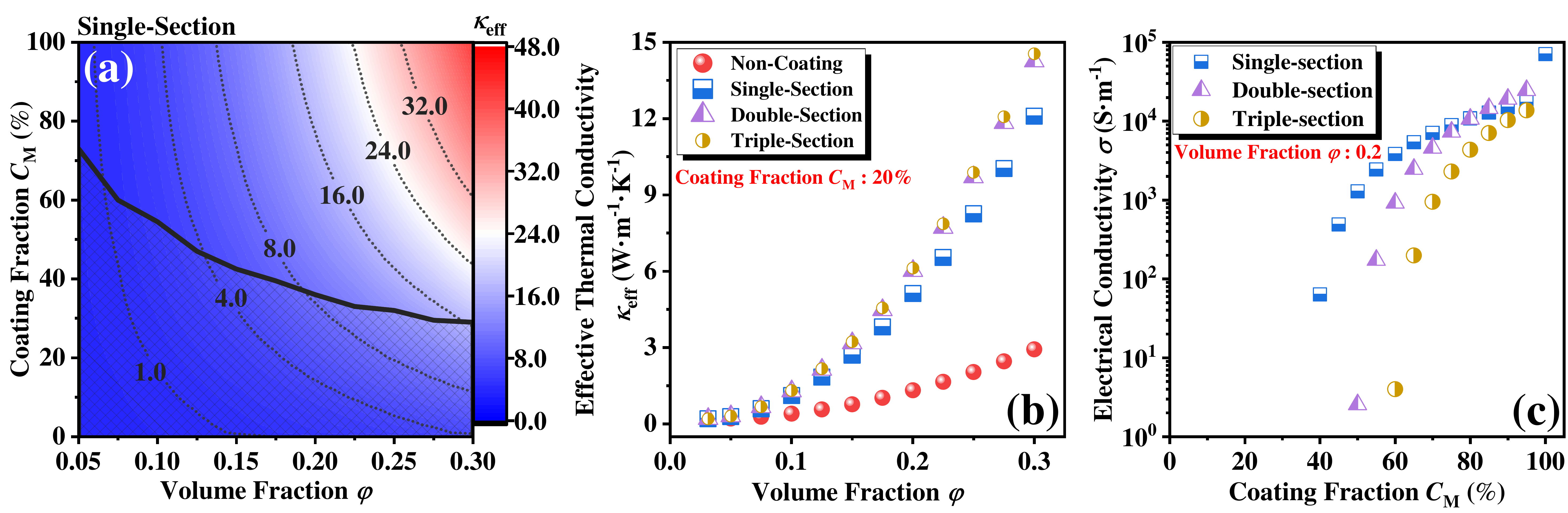}
	\caption{(Color online) (a) the calculated $\kappa_{\rm eff}$ as a function of $\varphi$ and $C_{\rm M}$ with the coated structure of one-section. The black line is the modified percolation threshold. (b) the calculated $\kappa_{\rm eff}$ as a variation of $\varphi$ and coated structure with the $C_{\rm M} = 20\%$. (c) the calculated $\sigma$ plotted as a function of $C_{\rm M}$ for different coating configurations when the $\varphi = 0.2$.} 
	\label{fig3}
\end{figure*}

\begin{figure*}[htp]
	\centering
	\includegraphics[width=0.9\linewidth]{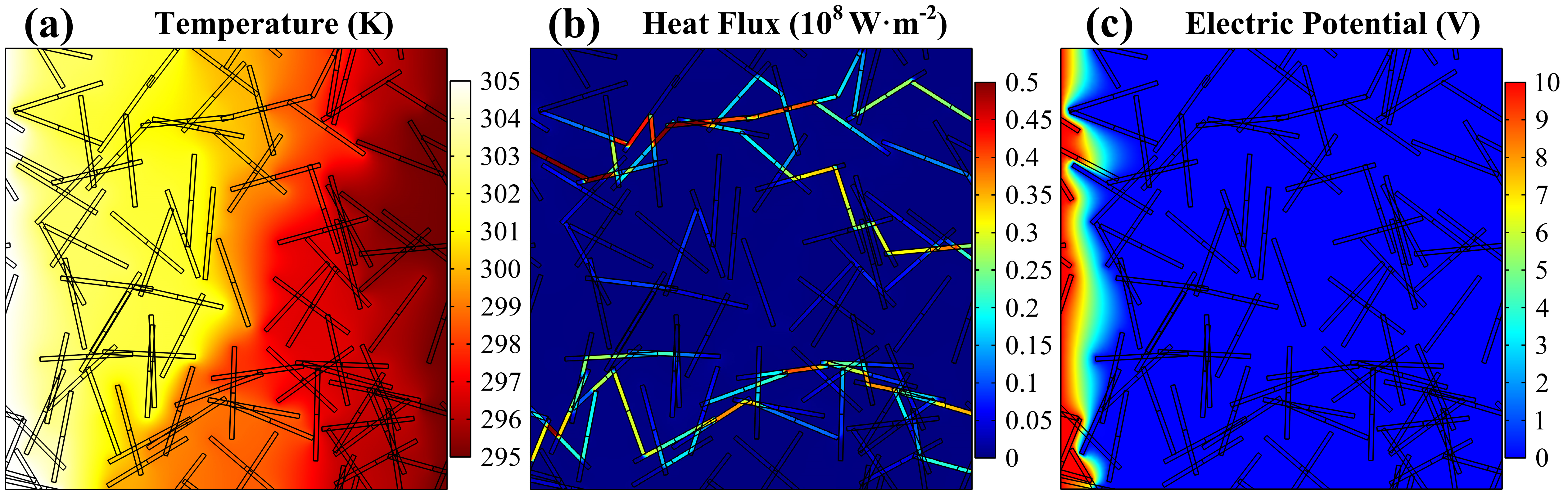}
	\caption{(Color online) The contour distribution of (a) temperature, (b) heat flux, and (c) electric potential with one-section, $C_{\rm M}=50\%$, and $\varphi=0.2$. The periodic boundary conditions are applied in the X and Y directions. The soft-core nanowires with $l/L=0.2$ are applied in 2D FEM simulation. The TBCs of Ag/PVDF and SiC/PVDF interface can be estimated by literature about 3$\times$10$^{7}$ and  1$\times$10$^{7}$\,W·m$^{-2}$·K$^{-1}$, respectively.\cite{monachon2016thermal,majumdar2004role,jin2011thermal} And the electrical conductivity of PVDF and 6H-SiC are set as 3.39$\times$10$^{-15}$ and 2$\times$10$^{-10}$\,S·m$^{-1}$, respectively.\cite{gradinaru1997electrical,wang2018tactile}}
	\label{fig4}
\end{figure*}

After validating the calculation method, we investigate the effective overall thermal conductivity of the composite materials. It depends on several parts: the thermal conductivity of matrix, the thermal percolating network, and the fillers which do not belong to the network (isolated fillers and finite clusters) \cite{huxtable2003interfacial,malekpour2014thermal,yu2008enhanced}. Here, we focus on the contribution of the thermal percolating network and the matrix to the effective thermal conductivity because the contribution of isolated fillers and finite clusters are trivial. 
Thus, the effective thermal conductivity is characterized by $\kappa_{\rm eff} \approx \kappa_{\rm net}+\kappa_{\rm p}$, where $\kappa_{\rm net}$ and $\kappa_{\rm p}$ are the thermal conductivity of thermal percolating network and matrix, respectively.
Assuming the matrix is thermal insulated, the thermal conductivity of thermal percolation network can be obtained by applying a temperature bias on the left and right planes of the system and then calculate the heat flux,  which can be obtained through a thermal resistor network algorithm and Fourier's law. %for given volume fractions. 
Due to the existence of TBRs, the effective thermal conductivity of partially coated nanowires $\kappa_{\rm cnw}$ are modified as
\begin{equation}
	\kappa_{\rm cnw}=\dfrac{4l\kappa_{\rm nw}/R_{\rm I}}{4l/R_{\rm I}+\pi\frac{d^{2}}{\delta}\kappa_{\rm nw}},
	\label{kappa}
\end{equation}
where $\kappa_{\rm nw}$ is the thermal conductivity of SiC nanowires, $\delta$ is chosen to be around $0.04d^{2}$, $R_{\rm I}$ is the average TBR between fillers which is evaluated by a weighted averaging of $R_{\text{Ag-Ag}}$, $R_{\text{Ag-SiC}}$ and $R_{\text{SiC-SiC}}$ according to the probability of these contacts counted in the generated structure. The value of $R_{\text{Ag-Ag}}$, $R_{\text{Ag-SiC}}$ and $R_{\text{SiC-SiC}}$ are taken as 2$\times$10$^{-10}$, 2$\times$10$^{-9}$, 1$\times$10$^{-8}$\,m$^{2}$·K·W$^{-1}$, respectively  \cite{gundrum2005thermal,swartz1989thermal,monachon2016thermal,zhong2021thermal}.  
The poly (vinylidene fluoride) (PVDF) was selected as the matrix material due to its high temperature tolerance and good thermal stability, and the thermal conductivity of PVDF $\kappa_{\rm p}$ is 0.19\,W·m$^{-1}$·K$^{-1}$ \cite{han2011thermal,xi2020ubiquitous}. 
The thermal conductivity of SiC is 490\,W·m$^{-1}$·K$^{-1}$ \cite{park1998sic}, and it is dielectric \cite{gradinaru1997electrical}. The aspect ratio of SiC nanowires is $l/d=20$ with $l=4\mu m$. The electrical conductivity of silver is 6.3$\times$10$^{7}$\,S·m$^{-1}$ \cite{wang2018cohesively}.

Figure \ref{fig3} (a) shows the calculated $\kappa_{\rm eff}$ as a function of the volume fraction $\varphi$ and coating fraction $C_{\rm M}$ with the coated structure of one-section. The threshold of electrical percolation is also plotted. 
We found that the composite materials with ultra-high thermal conductivity (about 8\,W·m$^{-1}$·K$^{-1}$) and electrical insulation could be achieved when $0.22 \leq \varphi \le 0.3$ and $C_{\rm M} = 30\%$.
%formation of electrical percolating networks can be regulated by the volume fraction $\varphi$, coating fraction $C_{M}$, and coating configurations.
%For one section configuration, the modified electrical percolation thresholds are enhanced as shown in Fig.\ref{fig3}(a).
And $\kappa_{\rm eff}$ can further be  improved by adding the number of sections $N$ when the coating fraction $C_{\rm M}$ is fixed to 0.2 as shown in Fig.\ref{fig3}(b). Moreover, coating configuration with larger number of sections results in a larger value of $C_{\rm M}$ which maintain the electrical insulation as shown in Fig.\ref{fig3}(c). This is helpful to achieve a larger value of $\kappa_{\rm eff}$.
As a result, our calculation results show that the concept of TP-EI with partially metallic coating is effective and successful.

In order to present an intuitive sense of our concept. We also used a 2D finite element model (FEM) to verify the successful of the TP-EI concept. Please note that the computational cost of 3D FEM calculation is too large to be obtained. We present the 2D case to conceptually demonstrate our idea which works both in 2D and 3D.  
Figure \ref{fig4} shows the contour distribution of temperature, heat flux, and electric potential with one-section, $C_{\rm M}=50\%$, and $\varphi=0.2$.
Two heat flux channels are clearly shown in the contour distribution of heat flux, while the electrical percolating networks do not form in the contour distribution of electric potential. 

In conclusion, we have proposed a type of composite materials where insulating nanowires with partially metallic coating are embedded in an insulating polymer. 
The thermal conductivity can be increased several times by coating the dielectric fillers with metal layer, due to the reduction of TBRs between fillers. 
Meanwhile, Electrical conductivity can also be regulated by $C_{\rm M}$ and coated structure. 
Our calculated results show that the concept of (TP-EI) with partially metallic coating structures is effective and successful to achieve efficient TIMs with ultra-high thermal conductivity and electrical insulation. 
Our findings are not only valid for the design of TIMs, but also applicable for other applications requiring low TBR. 
%\\

\begin{acknowledgments}
This work was supported by National Key R{\&}D Program of China
(No. 2017YFB0406004), National Natural Science Foundation of China
(No. 11890703), Key-Area Research and Development Program of Guangdong Province (No. 2020B010190004).
\end{acknowledgments}

\section*{Conflict of interest}
The authors have no conflicts to disclose.

\section*{DATA AVAILABILITY}
The data that support the findings of this study are available within the article.

\normalem
%\bibliography{ref}

%

\end{document}